\PassOptionsToPackage{table}{xcolor}
\documentclass[10pt,conference]{IEEEtran}

\usepackage{tabularx,booktabs,threeparttable,array}
\usepackage{pifont}
\usepackage{multicol,multirow}
\usepackage{xcolor}
\usepackage{graphicx}
\usepackage{listings}
\usepackage{caption}
\usepackage{url}
\usepackage[hidelinks]{hyperref}
\usepackage{amsmath} 

\newcolumntype{C}{>{\centering\arraybackslash}X}

\definecolor{codegreen}{rgb}{0,0.6,0}
\definecolor{codegray}{rgb}{0.5,0.5,0.5}
\definecolor{codepurple}{rgb}{0.58,0.3,0.82}
\definecolor{backcolour}{rgb}{0.98,0.99,0.98}
\definecolor{HardStatic}{RGB}{255, 237, 237}

\makeatletter
\newcommand{\fullbg}[2]{%
  \begingroup
  \setlength{\fboxsep}{0.1pt}%
  \colorbox{#1}{\strut\lst@basicstyle\mbox{#2}}%
  \endgroup
}

\newcommand{\hblue}[1]{\fullbg{blue!15}{#1}}
\makeatother

\newcommand{\pos}[1]{\cellcolor{green!15}\textbf{#1}}
\newcommand{\negcell}[1]{\cellcolor{red!15}#1}
\newcommand{\neu}[1]{\cellcolor{gray!12}#1}
\newcommand{\Fonehat}{\ensuremath{\widehat{F}_1}}

\lstdefinelanguage{yaml}{
  sensitive=false,
  comment=[l]{\#},
  morestring=[b]",
  morestring=[b]',
  keywords={true,false,null,yes,no,on,off}
}

\lstdefinestyle{mystyleSix}{
    style=mystyle,
    basicstyle=\ttfamily\fontsize{7}{9}\selectfont
}

\lstdefinestyle{mystyle}{
    language=python,
    backgroundcolor=\color{backcolour},
    commentstyle=\color{codegreen},
    keywordstyle=\color{magenta},
    numberstyle=\tiny\color{codegray},
    stringstyle=\color{codepurple},
    basicstyle=\ttfamily,
    columns=fullflexible,
    breaklines=true,
    breakatwhitespace=false,
    captionpos=b,
    keepspaces=true,
    numbers=left,
    numbersep=5pt,
    showspaces=false,
    showstringspaces=false,
    showtabs=false,
    tabsize=2,
    frame=tb,
    rulecolor=\color{gray},
    framerule=0.9pt,
    escapeinside={(*}{*)}
}

\usepackage[linesnumbered,ruled,vlined]{algorithm2e}
\usepackage{tcolorbox}
\tcbuselibrary{skins,breakable}
\newtcolorbox{rqtakeaway}[1]{%
  enhanced,
  breakable,
  colback=gray!6,
  colframe=gray!45,
  boxrule=0.35pt,
  arc=1pt,
  left=2pt,
  right=2pt,
  top=2pt,
  bottom=2pt,
  before skip=3pt,
  after skip=4pt,
  fonttitle=\bfseries,
  title={#1}
}
\newcommand{\rqheading}[1]{%
  \vspace{0.45em}%
  \noindent{\bfseries\itshape #1}\par%
  \vspace{0.15em}%
}
\newcommand{\github}[1]{\href{https://github.com/#1}{\texttt{#1}}}

\title{Beyond Local Patterns: Detecting Non-Local ML Code Smells with Code Property Graphs}

\author{%
\IEEEauthorblockN{
Brahim MAHMOUDI\IEEEauthorrefmark{1},
Naouel MOHA\IEEEauthorrefmark{1},
Quentin STI{\'E}VENART\IEEEauthorrefmark{2},
Florent AVELLANEDA\IEEEauthorrefmark{2}
}
\IEEEauthorblockA{
\IEEEauthorrefmark{1}{\'E}cole de technologie sup{\'e}rieure,
Montr{\'e}al, Qu{\'e}bec, Canada
}
\IEEEauthorblockA{
\IEEEauthorrefmark{2}Universit{\'e} du Qu{\'e}bec {\`a} Montr{\'e}al,
Montr{\'e}al, Qu{\'e}bec, Canada
}
}

\begin{document}
\maketitle

\begin{abstract}
Machine Learning (ML) pipelines encode quality-relevant decisions across data preparation, training, evaluation, and configuration code. Some recurring source-level quality problems in these pipelines, known as ML code smells, may not cause immediate failures but can harm reproducibility, robustness, efficiency, or maintainability. Detecting ML code smell occurrences is challenging because the decisive evidence is often non-local, spanning helper functions, wrappers, imports, control-flow, and data-flow relations.

We present \textit{SpecDetect4ML}, a static analyser that operationalises 22 ML code smells using CPG views with project-level resolution. We evaluate it on 890 Python ML-based systems comprising more than 20M LOC and a system-level recall benchmark over the complete ML-relevant source subset of 10 selected systems. Under identical ML code smell specifications, CPG-based reasoning raises recall from 68.62\% to 88.14\% compared with AST-only analysis, while keeping CPG precision comparable at 90.32\%. These results show that project-level static reasoning expands the detectable portion of non-local ML code smell occurrences, while configuration-dependent and runtime-only occurrences remain outside our source-only static claims.
\end{abstract}


\section{Introduction}
\label{sec:intro}

The rapid adoption of Machine Learning (ML) is reshaping how software systems are designed, developed, and maintained \cite{AsynthesisOfGreen2024,wei2022APIRecommendation,Shaw2022}. In this work, we focus on \textbf{ML-based systems}: Python systems that implement ML pipelines for data preparation, training, and evaluation. Such pipelines encode quality-critical decisions in source code, often through fast-evolving libraries and framework-specific conventions, which makes them difficult to reason about, test, and maintain \cite{passi2018,Zhang2022code,wei2022APIRecommendation}. We analyse Python source code that implements these pipelines, rather than learned model artefacts, deployed inference binaries, or complete AI-enabled systems around the pipeline.
A growing body of work studies \textbf{ML code smells}: recurring source-level patterns in ML pipeline code that may not be syntactically incorrect or immediately faulty, but can harm reproducibility, robustness, efficiency, or maintainability over time \cite{Zhang2022code,Fowler1999Refactoring}. Existing static analysers for ML code smells, including mlpylint \cite{Hamfelt2023MLpylint} and CodeSmile \cite{recupito2025_codesmile_lifecycle}, rely on Abstract Syntax Tree (AST)-based local pattern matching. An AST represents the syntactic structure of a program, such as calls, expressions, assignments, and statement nesting, and is effective when the decisive evidence is explicit in one call site or one file. However, AST matching alone does not encode how execution may flow between statements, how values are defined and reused, or how calls, imports, aliases, and helper functions connect across files.
To capture these relations, we use Code Property Graphs (CPGs)~\cite{Yamaguchi2014CPG}. A CPG enriches the AST with additional program-analysis graphs, including a Control-Flow Graph (CFG) for execution order, a Program Dependence Graph (PDG) for data and control dependencies, call-graph relations, and project-level links such as imports and cross-file references. Compared with a plain AST, a CPG therefore preserves both the local syntactic evidence and the non-local relations needed to reason about ordering, propagation, wrappers, aliases, and cross-file behaviour. This additional structure is important when an ML code smell occurrence depends on helper functions, wrappers, aliases, control flow, data flow, or cross-file relations.

\captionsetup[lstlisting]{
  labelfont=bf,
  textfont=bf
}

\lstset{
  aboveskip=1pt,
  belowskip=1pt,
  numbersep=3pt
}
{\ttfamily\scriptsize preprocess\_utils.py}
\begin{lstlisting}[
  style=mystyleSix,
  basicstyle=\ttfamily\scriptsize,
  language=python,
  numbers=left,
  stepnumber=1,
  firstnumber=1,
  escapeinside={(*}{*)}
]
from sklearn.preprocessing import StandardScaler as SS
def preprocess(X):
    return SS().fit_transform(X)
\end{lstlisting}

{\ttfamily\scriptsize data\_split.py}
\begin{lstlisting}[
  style=mystyleSix,
  basicstyle=\ttfamily\scriptsize,
  language=python,
  numbers=left,
  stepnumber=1,
  firstnumber=1,
  escapeinside={(*}{*)}
]
from sklearn.model_selection import train_test_split as tts
def split_dataset(X, y, seed):
    return tts(X, y, test_size=0.2, random_state=seed)
\end{lstlisting}\vspace{-4pt}

{\ttfamily\scriptsize train\_pipeline.py}\vspace{-1pt}
\begin{lstlisting}[
  style=mystyleSix,
  basicstyle=\ttfamily\scriptsize,
  language=python,
  numbers=left,
  stepnumber=1,
  firstnumber=1,
  escapeinside={(*}{*)}
]
import preprocess_utils as pu
from data_split import split_dataset
def run(cfg):
    X, y = load_dataset(cfg.path)
    if cfg.scale:
        (*\hblue{X = pu.preprocess(X)}*) # Cross-module call linking exposes the leakage
    Xtr, Xte, ytr, yte = split_dataset(X, y, cfg.seed)
    model.fit(Xtr, ytr)
    score = model.score(Xte, yte)
    return score
\end{lstlisting}
\captionof{lstlisting}{A non-local R11 \textit{Data Leakage} occurrence recoverable with CPG-based project-level resolution but not with AST-only analysis}
\label{lst:R11_cpg_recovered}

Listing~\ref{lst:R11_cpg_recovered} illustrates the limitation. R11 \textit{Data Leakage} occurs when preprocessing is applied to the full dataset before the training/test split, allowing test-set information to influence training and potentially leading to overly optimistic evaluations \cite{Zhang2022code,wei2024Demystifying}. In the example, scaling is implemented in \texttt{preprocess\_utils.py}, invoked through a module alias in \texttt{train\_pipeline.py}, and followed by a split defined in \texttt{data\_split.py}. Recovering this ML code smell occurrence therefore requires linking information across files rather than matching a single local AST pattern.
To address this limitation, we propose \textit{SpecDetect4ML}, a specification-driven static analyser for ML code smells. SpecDetect4ML separates smell specification from project-level resolution: executable specifications capture the 22 ML code smells from Zhang et al.'s catalogue \cite{Zhang2022code}, while CPG views combine AST, control-flow, data-flow, and cross-file links to recover source-level evidence that local AST matching misses. This gives the backend a concrete advantage on non-local cases because it preserves ordering constraints, connects derived values to later checks, and recovers evidence hidden behind wrappers, aliases, and helper abstractions. We target \emph{statically observable ML code smell occurrences}. We do not claim complete behavioural detection, and configuration-dependent or runtime-only cases remain outside our static claims.
We evaluate \textit{SpecDetect4ML} on 890 ML-based systems totalling over 20M lines of code. It achieves 90.32\% precision on alert audits and 88.14\% recall on a system-level benchmark inspected by six ML-experienced annotators. More importantly, the goal of the study is to isolate what project-level CPG reasoning contributes when the ML code smell specifications are held constant. A controlled AST-only/CPG comparison under identical ML code smell specifications therefore shows that CPG-based project-level resolution improves the recovery of statically observable non-local occurrences while keeping precision comparable. Our contributions are: (i) an occurrence-level detectability taxonomy for ML code smells, (ii) empirical evidence that CPG-based project-level resolution improves recovery over AST-only analysis, and (iii) a reusable implementation artefact for large-scale ML code smell detection.
The rest of this paper is organised as follows. Section~\ref{sec:smells_overview} overviews the ML code smell catalogue, Section~\ref{sec:spec4ai_methodology} details the approach and detectability taxonomy, Section~\ref{sec:resuls_and_eval} reports the validation design and results, Section~\ref{sec:related_work} discusses related work, and Section~\ref{sec:conclusion} concludes.

\section{ML Code Smells Overview}
\label{sec:smells_overview}
ML code smells are recurring source-level patterns in ML pipeline code that are not necessarily syntax errors, but signal risks for reproducibility, correctness, efficiency, or maintainability. A training script may run successfully while still leaking test information into preprocessing, omitting deterministic options, or relying on implicit data schema assumptions. In this paper, detections are therefore interpreted as \emph{review candidates}: source locations that deserve inspection before the code is treated as defective.
We build on Zhang et al.'s catalogue of 22 ML code smells~\cite{Zhang2022code}, an established reference catalogue with existing analysers and datasets. This fixed catalogue supports comparison against existing ML-smell tools without claiming exhaustive coverage of emerging ML engineering issues. Retaining all 22 smells, rather than only graph-friendly ones, also provides local occurrences for the AST-only comparison. Table~\ref{tab:smells_reading_map} groups the original identifiers from Zhang et al.'s catalogue by the ML pipeline aspect they primarily affect. This first column is a reading aid, not a new taxonomy or severity ranking. Full definitions, examples, recommended fixes, pipeline stages, and executable specifications are provided in the replication package~\cite{SpecDetect4MLReplication}.
\begin{table*}[t]
\centering
\scriptsize
\setlength{\tabcolsep}{1.8pt}
\renewcommand{\arraystretch}{0.84}
\caption{Reading map for Zhang et al.'s 22 ML code smells used in this study.}
\label{tab:smells_reading_map}
\newcommand{\rid}[1]{\texttt{\textbf{#1}}}
\newcommand{\sname}[1]{\textit{#1}}
\begin{tabularx}{\textwidth}{@{}>{\raggedright\arraybackslash}p{1.95cm}>{\raggedright\arraybackslash}p{0.74cm}>{\raggedright\arraybackslash}p{5.65cm}>{\raggedright\arraybackslash}X@{}}
\toprule
\rowcolor{gray!14}
\textbf{Pipeline aspect} & \textbf{ID} & \textbf{Smell name} & \textbf{Intuition} \\
\midrule
\textbf{Reproducibility} & \rid{R2} & \sname{Randomness Uncontrolled} &
Choices that make ML runs hard to reproduce, compare, or debug. \\
 & \rid{R5} & \sname{Hyperparameter Not Explicitly Set} & \\
 & \rid{R6} & \sname{Deterministic Algorithm Option Not Used} & \\
\midrule
\textbf{Training} & \rid{R4} & \sname{Improper Train/Eval Mode Toggling} &
Patterns that affect how training or evaluation code executes. \\
 \textbf{execution}& \rid{R8} & \sname{PyTorch Call Method Misused} & \\
 & \rid{R9} & \sname{Gradients Not Cleared} & \\
 & \rid{R10} & \sname{Memory Not Freed} & \\
\midrule
\textbf{Data} & \rid{R7} & \sname{Missing Mask of Invalid Value} &
Data-cleaning and feature-engineering choices that can corrupt, misrepresent,  \\
 \textbf{preparation}& \rid{R12} & \sname{Matrix Multiplication API Misused} & or destabilize downstream data.\\
 & \rid{R13} & \sname{Empty Column Misinitialization} & \\
 & \rid{R14} & \sname{DataFrame Conversion API Misused} & \\
 & \rid{R15} & \sname{Merge API Parameter Not Explicitly Set} & \\
 & \rid{R16} & \sname{In-Place APIs Misused} & \\
 & \rid{R18} & \sname{NaN Equivalence Comparison Misused} & \\
 & \rid{R20} & \sname{Chain Indexing} & \\
 & \rid{R21} & \sname{Columns and DataType Not Explicitly Set} & \\
 & \rid{R22} & \sname{No Scaling Before Sensitive Operation} & \\
\midrule
\textbf{Evaluation} & \rid{R11} & \sname{Data Leakage} &
Patterns that can make reported model quality misleading. \\
 \textbf{reporting}& \rid{R19} & \sname{Threshold-Dependent Validation} & \\
\midrule
\textbf{Runtime} & \rid{R1} & \sname{Broadcasting Feature Not Used} &
Missed library idioms that may waste memory or computation. \\
\textbf{efficiency} & \rid{R3} & \sname{TensorArray Not Used} & \\
 & \rid{R17} & \sname{Unnecessary Iteration} & \\
\bottomrule
\end{tabularx}
\end{table*}

This grouping also helps interpret static detectability. Some ML code smell occurrences expose decisive information in a single API call, while others require reasoning about statement order, helper functions, cross-file calls, data dependencies, configuration, or runtime behaviour. Section~\ref{subsec:taxonomy} makes this distinction explicit through the detectability taxonomy used throughout the evaluation.

\section{\textit{SpecDetect4ML}: Specification-Driven Static Analysis}
\label{sec:spec4ai_methodology}
\newcommand{\stepfield}[1]{\noindent{\bfseries #1:}}

\begin{figure}[htbp]
    \centering
    \includegraphics[width=0.72\linewidth]{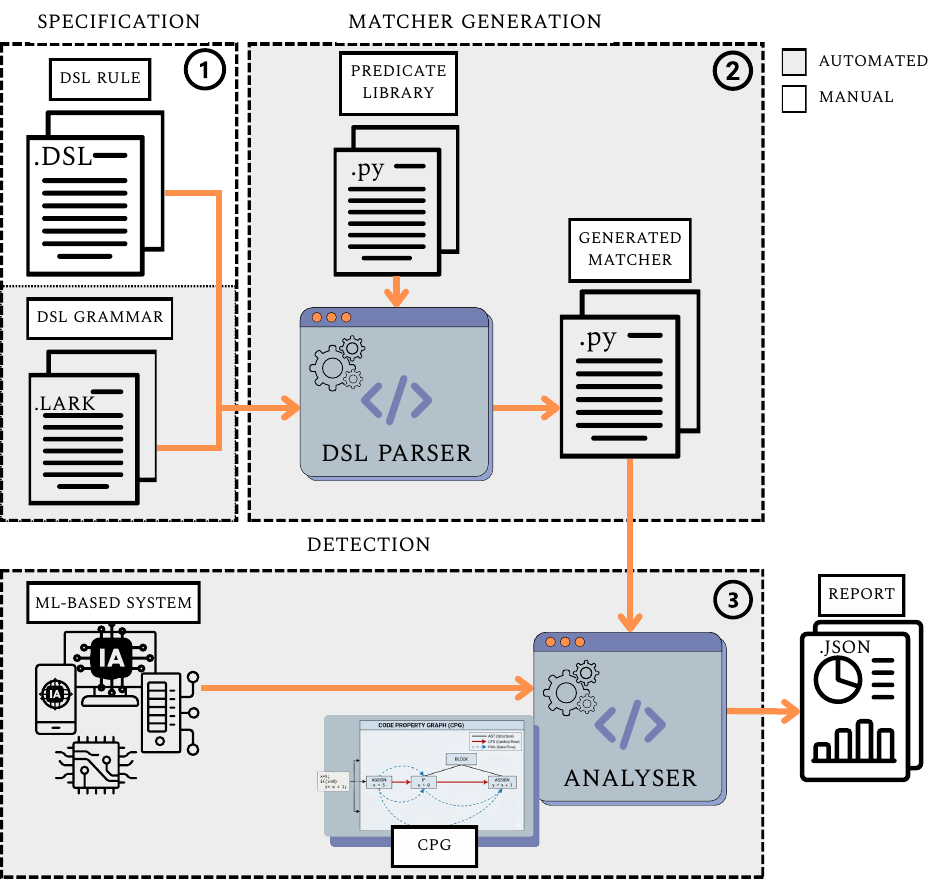}
    \caption{Overview of \textit{SpecDetect4ML}'s three step workflow.}
    \label{fig:overview_spec4ai}
\end{figure}

\textit{SpecDetect4ML} encodes ML code smell descriptions as executable detection rules evaluated with alternative static-analysis backends. Figure~\ref{fig:overview_spec4ai} presents the workflow. Rules are written in a declarative Domain-Specific Language (DSL) (Step~1), compiled into executable matchers (Step~2), and evaluated over project-level CPG views that expose control-flow and data-flow relations (Step~3). We describe our approach using R5 \textit{Hyperparameter Not Explicitly Set} as a running example.

\subsection{Detectability Taxonomy}
\label{subsec:taxonomy}

To make the scope of our claims explicit, we introduce a detectability taxonomy that classifies \emph{ML code smell occurrences} according to where the decisive evidence resides and which form of static reasoning is needed to recover it. By \emph{decisive evidence}, we mean the source-level facts needed to decide whether a reported location constitutes an ML code smell occurrence under the catalogue definition. Such evidence may be a single API call, an omitted argument, the relative order of preprocessing and splitting operations, a def-use relation, or a call chain spanning multiple files. We derived the taxonomy from Zhang et al.'s smell definitions and pilot inspection of validation cases by separating evidence along two axes: where it appears in the project and whether it is statically observable. The taxonomy distinguishes occurrences recoverable by local AST matching, occurrences that benefit from project-level CPG reasoning, and cases whose decisive evidence depends on configuration or runtime behaviour. It is therefore used throughout the evaluation to interpret what is recovered, what is missed, and why.

\begin{itemize}
\item \textbf{T1: \textit{Local API-level occurrences}:} the decisive evidence is syntactically explicit in a single statement or call site.
\item \textbf{T2: \textit{Intra-file ordering occurrences}:} the decisive evidence depends on statement order or local control flow within one file.
\item \textbf{T3: \textit{Same-file interprocedural occurrences}:} the decisive evidence is distributed across multiple functions within the same file.
\item \textbf{T4: \textit{Cross-file occurrences}:} the decisive evidence is distributed across files and requires linking imports, aliases, wrappers, or calls across modules.
\item \textbf{T5: \textit{Configuration-dependent occurrences}:} the decisive evidence depends on values supplied through configuration files, command-line arguments, or environment settings.
\item \textbf{T6: \textit{Runtime-only occurrences}:} the decisive evidence becomes visible only through runtime behaviour, reflective execution, or environment-dependent effects.
\end{itemize}

AST-only analysis mainly covers T1 and part of T2. CPG-based project-level reasoning covers these cases and further supports T3 and T4 through control-flow, data-flow, same-file interprocedural, and cross-file relations. T5 is only partially recoverable when configuration values are statically visible in the inspected source code, while T6 remains outside the capabilities of static analysis. The taxonomy is occurrence-level rather than smell-level: the same ML code smell may be local in one project, non-local in another, or configuration-dependent when assembled externally, as with R11 \textit{Data Leakage}. We use the taxonomy to interpret concrete successes and failures, not to assign fixed detectability classes to ML code smells.

\subsection{Step 1: Specification}
\label{sub:Step1}

\stepfield{Input} A textual description of a ML code smell.

\stepfield{Output} A declarative DSL rule $r$ that specifies detection criteria through high-level predicates.

\stepfield{Implementation} We manually specify each smell description as a declarative detection rule in our DSL, where each rule corresponds to a quantified first-order formula composed of predicates. We provide 87 reusable predicates that abstract over low-level syntactic details into ML-oriented properties~\cite{SpecDetect4MLReplication}. Predicates are referenced in a representation-agnostic manner, so rules state \emph{what} property must hold without encoding \emph{how} it is computed. We define the DSL in Lark’s Extended Backus-Naur Form (EBNF)~\cite{Wirth1977EBNF,lark}. Listing~\ref{fig:grammar} shows a simplified excerpt of the grammar.

\begin{lstlisting}[style=mystyle,basicstyle=\ttfamily\scriptsize, language=python, morekeywords={fit,train_test_split}, caption=Excerpt of the \textit{SpecDetect4ML} DSL grammar, label = fig:grammar]
rule_definition: "rule" IDENTIFIER STRING ":" 
                 "condition:" cond_expr 
                 "action:" action_expr
cond_expr: "exists" IDENTIFIER "in" "CPG" ":" "(" cond_expr ")"
              | predicate_call
              | "not" cond_expr
              | cond_expr "and" cond_expr
              | cond_expr "or" cond_expr
              | "(" cond_expr ")"
predicate_call: IDENTIFIER "(" argument_list? ")"
action_expr: "report" STRING
\end{lstlisting}

\stepfield{Running Example} Listing~\ref{fig:RuleR5} is the DSL rule for R5 \textit{Hyperparameter Not Explicitly Set}. The rule states that a smell occurrence exists when an ML method call (\texttt{isMLMethodCall}) is detected without explicitly specified hyperparameters (\texttt{hasExplicitHyperparameters}).

\begin{lstlisting}[
style=mystyle,
basicstyle=\ttfamily\scriptsize,
alsoletter={_},
caption={DSL definition of R5},
label={fig:RuleR5},
emph={rule,condition,action,report},
emphstyle={\color{magenta}},
emph={[2]exists,isMLMethodCall,hasExplicitHyperparameters,in,CPG},
emphstyle={[2]\color{red}}
]
rule R5 "Hyperparameter Not Explicitly Set":
   condition:
        exists call in CPG: (
           isMLMethodCall(call) and not
           hasExplicitHyperparameters(call) )
   action: report "Hyperparameter not explicitly set at line{lineno}"
\end{lstlisting}

The DSL is a specification layer rather than the main source of the detection gains reported in this paper. General-purpose frameworks such as CodeQL~\cite{codeql} and Semgrep~\cite{semgrep} can encode local and syntactic ML code smell checks, and the replication package includes representative R5 \textit{Hyperparameter Not Explicitly Set} specifications in both frameworks. The limitation for our setting is not raw expressiveness, but ML-specific reuse and controlled comparison: alias normalisation, helper resolution, and repeated library checks would need to be re-encoded across rules, which makes like-for-like AST/CPG ablation harder to maintain. AST-only and CPG modes share the same rule texts, matcher generator, predicate signatures, and occurrence-report format, while differing only in the relations that predicates may consult. The DSL therefore exposes 87 ML-oriented predicates, hides repeated framework-specific checks, and decouples ML code smell specifications from the analysis backend.

\noindent\textbf{Specification-authoring check.}
We use the DSL to support reuse across related ML checks. As an exploratory authoring check, five industry ML practitioners added two additional specifications: R23 \textit{EarlyStopping Not Used in Model.fit}~\cite{KerasEarlyStopping} with one new predicate, and R24 \textit{Index Column Not Explicitly Set in DataFrame Read}~\cite{pandasReadCSV} using existing predicates. Both are released as authoring artefacts and excluded from the 22-smell detection evaluation. We use this exercise only to check that the workflow can express additional smells with limited predicate reuse, not as general usability evidence.

\subsection{Step 2: Matcher Generation}
\label{sub:Step2}

\stepfield{Input} A DSL rule $r$ specified in Step 1, the DSL grammar $\mathcal{G}$, and the predicate library interface $\mathcal{P}$.

\stepfield{Output} A generated Python matcher file $m$ that implements the rule logic as compiled predicate calls and control structures.

\stepfield{Implementation} Our parser takes as input $r$ and $\mathcal{G}$ to produce a rule parse tree, validates that all referenced predicates are defined in $\mathcal{P}$ and used consistently with their signatures, compiles the rule condition into a boolean expression over predicate calls, and emits a Python matcher file whose predicate evaluation is deferred to Step 3.
\begin{algorithm}[htbp]
\scriptsize
\caption{Matcher Generation Process For a Rule Rx}
\label{algo:hyperparam_rule}
\KwIn{$r_{Rx}$: DSL rule text, $\mathcal{G}$: DSL grammar, $\mathcal{P}$: predicate signatures}
\KwOut{$m_{Rx}$: generated Python matcher file implementing $r_{Rx}$}
$T \leftarrow \textsc{Parse}(r_{Rx}, \mathcal{G})$\;
$\textsc{Validate}(T, \mathcal{P})$\;
$E \leftarrow \textsc{Compile}(T)$ \tcp*[r]{boolean expression of predicate calls}
$m_{Rx} \leftarrow \textsc{EmitPython}(E)$\;
\Return{$m_{Rx}$}\;
\end{algorithm}

\subsection{Step 3: Detection}
\label{sub:Step3}

\stepfield{Input} The generated matcher files $m$ from Step 2 and the source code of an ML-based system $\mathcal{S}$.

\stepfield{Output} A structured JSON report $J$ that records detected smell occurrences with their rule identifiers, source locations, and information.

\stepfield{Implementation} As summarised in Algorithm~\ref{algo:step3_refined_linked_nod}, Step~3 evaluates the generated matchers over a static representation of $\mathcal{S}$. Each Python file is parsed into an AST, while \texttt{ProjectIndex} resolves statically visible project-local definitions, aliases, imports, wrappers, helper calls, and relative imports when their targets are present. It does not attempt sound treatment of star or dynamic imports, reflective execution, \texttt{eval}/\texttt{exec}, monkey patching, inheritance-dependent dispatch, or late runtime rebinding. We call this project-level reasoning: file-level CPG views derived from AST, CFG, and PDG edges are connected through \texttt{ProjectIndex}, rather than assembled into a monolithic whole-program graph. AST-only mode restricts predicates to local AST evidence, lexical statement order, and file-local syntax. It does not use resolved imports, call graphs, or def-use links. CPG mode lets the same predicates additionally consult CFG, PDG, and resolved project links. Propagation is bounded by $D$, the project maximum acyclic call-chain depth in \texttt{ProjectIndex}. Cycles are detected and broken during traversal.
\begin{algorithm}[htbp]
\scriptsize
\caption{Step 3 Detection}
\label{algo:step3_refined_linked_nod}
\KwIn{$\mathcal{S}$: ML-based system, $m$: generated matcher files}
\KwOut{$J$: JSON report of detected rule violations}
$F \leftarrow$ all \texttt{.py} files in $\mathcal{S}$\;
$\mathcal{R} \leftarrow \texttt{LoadMatchers}(m)$\;
\ForEach{$p \in F$}{
  $A[p] \leftarrow \texttt{ParseAST}(p)$\;
}
$I \leftarrow \texttt{BuildProjectIndex}(A)$\;
$D \leftarrow \texttt{MaxAcyclicCallDepth}(I)$\;
\ForEach{$p \in F$}{
  $CFG[p] \leftarrow \texttt{BuildCFG}(A[p])$\;
  $PDG[p] \leftarrow \texttt{BuildPDG}(A[p], CFG[p])$\;
  $RC[p] \leftarrow \texttt{ResolveCalls}(A[p], p, I)$\;
  $G[p] \leftarrow \texttt{AssembleFileCPG}(A[p], CFG[p], PDG[p])$\;
  $V[p] \leftarrow \texttt{FileCPGView}(p, G[p], I, RC[p], D)$\;
}
$\texttt{findings} \leftarrow [\ ]$\;
\ForEach{$r \in \mathcal{R}$}{
  \ForEach{$p \in F$}{
    $\texttt{findings} \leftarrow \texttt{MergeFindings}(\texttt{findings}, r.\texttt{match}(V[p]))$\;
  }
}
$J \leftarrow \texttt{SerialiseAsJSON}(\texttt{findings})$\;
\texttt{WriteFile}$(J)$\;
\Return{$J$}\;
\end{algorithm}

\noindent\textbf{Scope and limitations.}
\textit{SpecDetect4ML} targets statically observable ML code smell occurrences in Python source code. Its current backend is Python-specific, but the DSL is designed as a specification layer above the analysis backend: supporting another programming language would require implementing the corresponding parsing, CPG construction, and predicate semantics, while preserving the high-level ML code smell specifications when their meaning remains language-independent. The CPG-based backend extends AST-only analysis with project-level reasoning over control-flow, data-flow, and cross-file relations, improving recovery of non-local occurrences whose decisive evidence remains visible in the inspected source code. Occurrences whose decisive evidence depends on runtime behaviour or external configuration remain outside our static claims, especially under reflective constructs, \texttt{eval} or \texttt{exec}, highly dynamic imports, monkey patching, or late binding. Dynamic analysis could recover some of this missing evidence, but it requires executable environments, representative inputs, and additional runtime cost. We therefore use project-level static analysis as a practical compromise: it expands recovery of statically observable non-local occurrences while preserving scalability across large and heterogeneous repositories.

\section{Validation Design and Results}
\label{sec:resuls_and_eval}

In this section, we present the study used to validate our approach. We follow a mixed-method validation design combining quantitative measurements with qualitative inspection of detected ML code smell occurrences.

\subsection{Research Questions}
\label{sec:research_questions}

We answer five research questions.

\noindent\textbf{RQ$_1$.} What portion of the ML code smell catalogue is statically observable at scale, and how are the resulting occurrences distributed across smells, files, and systems?\par

\noindent\textbf{RQ$_2$.} How much does project-level CPG reasoning improve detection over AST-only analysis under identical ML code smell specifications and validation protocol?\par

\noindent\textbf{RQ$_3$.} How effective is \textit{SpecDetect4ML}, in precision, recall, and \Fonehat{}, compared with published static analysers for ML code smells and normalised LLM baselines?\par

\noindent\textbf{RQ$_4$.} What do representative true positives, false positives, and false negatives reveal about statically observable, configuration-dependent, and runtime-only ML code smell occurrences?\par

\noindent\textbf{RQ$_5$.} To what extent does \textit{SpecDetect4ML} scale to a large corpus of ML-based systems, and what overhead does project-level CPG reasoning introduce over AST-only analysis?

\subsection{Analysers and Baselines}
\label{subsec:analysers_baselines}

We validate \textit{SpecDetect4ML} against two published Python static analysers for ML code smells, mlpylint~\cite{Hamfelt2023MLpylint} and CodeSmile~\cite{recupito2025_codesmile_lifecycle}, and three LLM baselines, Qwen 3 30B Instruct, Claude Sonnet 4.6, and DeepSeek V4 Flash. Comparability is enforced through a shared occurrence-level protocol rather than through a shared implementation substrate: all external outputs are normalised to the same detection unit and matched against the same adjudicated benchmark.

CodeSmile covers 12 ML code smells from Zhang et al.'s catalogue~\cite{Zhang2022code}, so we compute its results on the corresponding 12-smell intersection using the adjudicated ground truth described in Section~\ref{subsec:gt}. mlpylint overlaps with 20 of the 22 ML code smells and is evaluated on the corresponding 20-smell intersection under the same occurrence-level protocol. The three LLM baselines are rerun across the full set of 22 ML code smell specifications under the same matching protocol. We selected one locally runnable instruction model, Qwen 3 30B Instruct, to support reproducibility without external service dependencies, one frontier service model, Claude Sonnet 4.6, and one paid low-latency model, DeepSeek V4 Flash, to represent practical deployment choices under realistic cost constraints. The exact provider model identifiers, access dates, prompts, temperatures, and response-normalisation scripts are recorded in the replication package. The goal is not to exhaustively benchmark all possible LLMs, but to compare \textit{SpecDetect4ML} with representative static and prompt-based LLM alternatives under a common, auditable occurrence-level protocol.

\subsection{Corpus Construction}
\label{subsec:corpus}
We built the 890-system corpus from GitHub repositories likely to contain Python ML pipeline code, using a three-stage procedure: query-based retrieval, metadata screening, and source-level screening.
\noindent\textit{Query-based retrieval.}
Candidate repositories were collected through GitHub search using broad query families rather than a single framework-specific query. Queries required Python as the primary language and combined ML-task terms, framework/library signals, and pipeline-action terms. The retrieval schema was:

\begin{tcolorbox}[colback=gray!6, colframe=gray!40, arc=1pt, boxrule=0.3pt, left=2pt, right=2pt, top=2pt, bottom=2pt]
\scriptsize
\raggedright
(\textcolor{red!65!black}{``machine learning''} \textcolor{blue!60!black}{OR} \textcolor{red!65!black}{``deep learning''} \textcolor{blue!60!black}{OR} \textcolor{red!65!black}{``data science''}) 
\textcolor{blue!60!black}{AND}\\
(\textcolor{red!65!black}{``TensorFlow''} \textcolor{blue!60!black}{OR} \textcolor{red!65!black}{``PyTorch''} \textcolor{blue!60!black}{OR} \textcolor{red!65!black}{``scikit-learn''} \textcolor{blue!60!black}{OR} \textcolor{red!65!black}{``Keras''} \textcolor{blue!60!black}{OR} \textcolor{red!65!black}{``XGBoost''}) 
\textcolor{blue!60!black}{AND}\\
(\textcolor{red!65!black}{``training''} \textcolor{blue!60!black}{OR} \textcolor{red!65!black}{``pipeline''} \textcolor{blue!60!black}{OR} \textcolor{red!65!black}{``experiment''} \textcolor{blue!60!black}{OR} \textcolor{red!65!black}{``evaluation''})
\end{tcolorbox}

\noindent\textit{Metadata and source screening.}
After duplicate removal, we retained repositories whose metadata, dependency manifests, or inspected Python source indicated analysable ML pipeline code. Retention required evidence of at least one ML-library signal, such as TensorFlow, PyTorch, scikit-learn, Keras, or XGBoost, and at least one identifiable pipeline stage, such as data preparation, model definition or training, evaluation, or experiment configuration. This screening excluded repositories whose ML relevance was only nominal or whose source did not provide Python code suitable for static analysis.
\noindent\textit{Recorded metadata.}
For each retained repository, the corpus records owner/name, description, star and fork counts, open issue count, creation and last modification dates, primary language, detected ML-library signals, test presence, and size proxies. The replication package~\cite{SpecDetect4MLReplication} releases this metadata together with the sampling scripts used to select the precision and recall validation subsets.

\subsection{Sampling Design and Ground Truth Construction}
\label{subsec:gt}

\begin{figure}[htbp]
    \centering
    \includegraphics[width=0.9\linewidth]{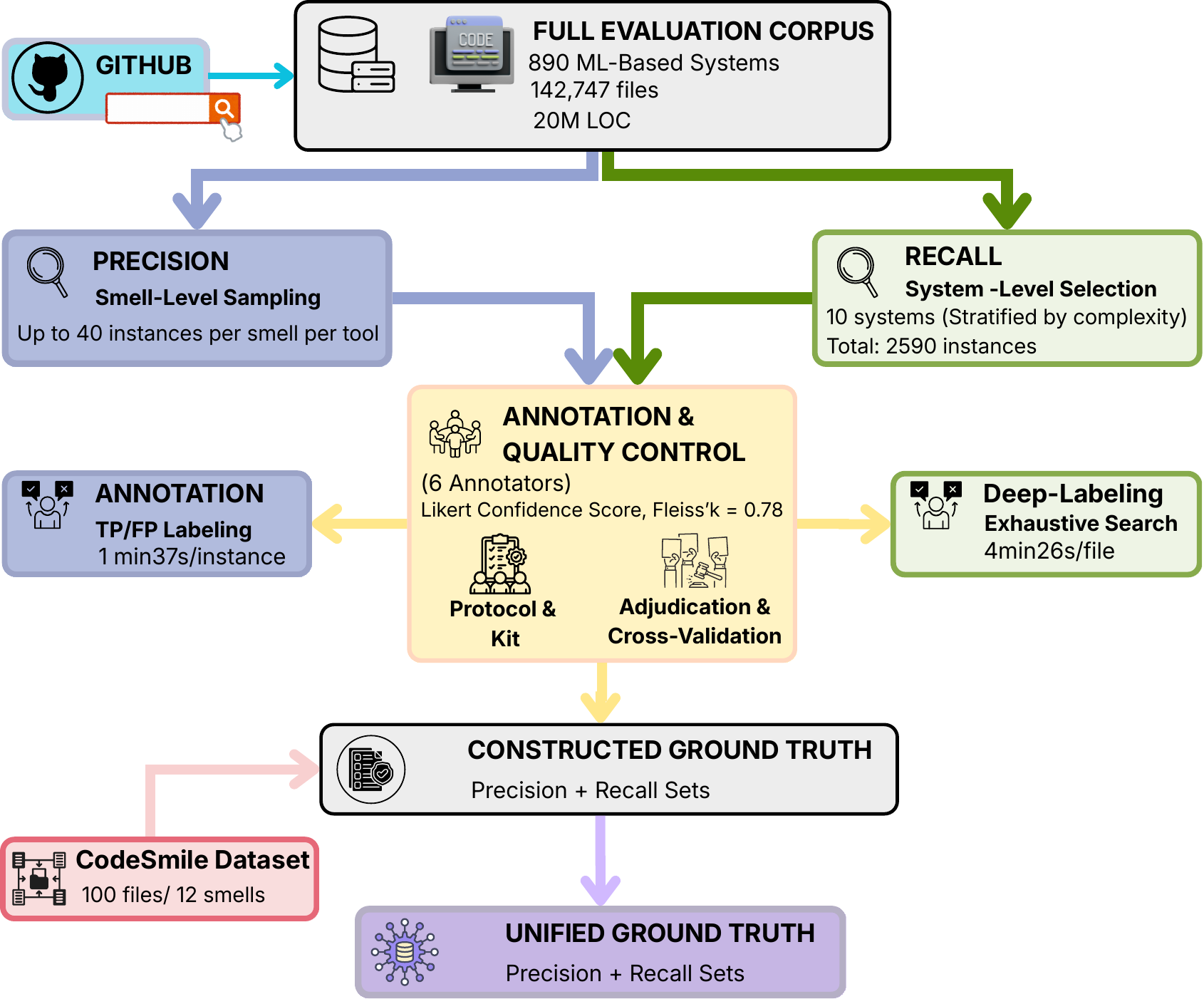}
    \caption{Overview of ground truth construction.}
    \label{fig:overview_gt}
\end{figure}

\begin{table}[t]
\centering
\caption{Selected systems for recall ground-truth construction. The \textbf{Total .py Files} column reports all Python files in the selected repositories before ML-relevance filtering.}
\label{tab:gt-recall-sample}
\scriptsize
\setlength{\tabcolsep}{1.6pt}
\renewcommand{\arraystretch}{0.65}
\begin{tabularx}{\columnwidth}{@{}X r r r l r l@{}}
\toprule
\textbf{System} & \textbf{Total .py Files} & \textbf{LLOC} & \textbf{Avg. CC} & \textbf{Bin} & \textbf{Contribs} & \textbf{Tests} \\
\midrule
\github{mindorii/kws} & 3 & 245 & 2.1 & low & 3 & No \\
\github{balakg/posewarp-cvpr2018} & 3 & 412 & 3.4 & low & 1 & No \\
\github{edouardoyallon/pyscatwave} & 7 & 1,204 & 2.8 & low & 14 & Yes \\
\github{tomgoldstein/loss-landscape} & 13 & 2,840 & 4.1 & mid & 8 & No \\
\github{google-deepmind/dm\_control} & 23 & 5,120 & 3.9 & mid & 42 & No \\
\github{alteryx/featuretools} & 22 & 8,450 & 4.5 & mid & 180 & Yes \\
\github{ixaxaar/pytorch-dnc} & 19 & 4,100 & 5.2 & mid & 5 & Yes \\
\github{microsoft/nni} & 373 & 112,400 & 5.8 & high & 350 & Yes \\
\github{dmlc/dgl} & 806 & 245,000 & 6.2 & high & 410 & Yes \\
\href{https://github.com/google-research/google-research}{\texttt{google-research}} & 2,586 & 890,500 & 5.5 & high & 1,500+ & Yes \\
\bottomrule
\end{tabularx}
\end{table}

The full evaluation corpus contains 890 ML-based systems, 142,747 Python files, and more than 20M lines of code. Building an exhaustive occurrence-level ground truth at this scale would require thousands of human-hours. We therefore use a two-track validation design, shown in Figure~\ref{fig:overview_gt}, that separates alert validation from missed-occurrence estimation. The precision track starts from analyser outputs and asks whether sampled alerts correspond to valid ML code smell occurrences. The recall track starts from independent manual inspection and asks which adjudicated occurrences each analyser recovers. This separation avoids circularity because precision labels are derived from sampled alerts, whereas recall labels are constructed before matching analyser outputs back to the benchmark.

For precision, we construct a smell-balanced audit sample over the complete corpus. For each ML code smell, we sample at least 40 distinct alerts when available. If fewer than 40 alerts are emitted for a smell, all available alerts are inspected. Each sampled alert is labelled as a true positive or a false positive against the smell-specific guideline, yielding smell-level precision estimates with 95\% Wilson confidence intervals. Aggregate precision pools inspected alerts and does not impute labels for uninspected warnings.

For recall, we construct a stratified system-level benchmark from the corpus. We treat logical lines of code as the primary size indicator and stratify systems using a composite proxy that jointly captures size and structural complexity through logical density, average cyclomatic complexity, and call-graph connectivity. We then select 10 systems across low, mid, and high size-complexity bins, as reported in Table~\ref{tab:gt-recall-sample}. The selected systems range from 245 to over 890,000 LLOC, include repositories with and without tests, and cover compact experimental repositories, medium-sized ML libraries, and large modular frameworks. Together, they contain 3,855 Python files before ML-relevance filtering.

The recall benchmark is system-level rather than alert-level. Six ML-experienced annotators inspected the complete ML-relevant source subset of the 10 selected systems, as defined by our ML-relevance filtering protocol, yielding 605 Python files and 2,547 adjudicated occurrence-level decisions. This trades corpus breadth for independent ground-truth depth: recall comes from inspected source code, while the 890-system corpus supports prevalence and scalability analysis.

The ML-relevance filtering protocol retains Python files related to model construction, training, evaluation, data preprocessing, inference, hyperparameter configuration, or ML pipeline logic, and excludes packaging, documentation, generic utilities, wrappers, tests without ML logic, and infrastructure code. The 605 files are therefore the complete ML-relevant source subset of the selected systems, not the complete set of Python files in those repositories. Recall is measured over known smell occurrences in this independently annotated benchmark, not as an exhaustive estimate of all missed occurrences in the full 890-system corpus.

To reduce implementation bias, the initial ground truth was constructed by non-author annotators who were blind to \textit{SpecDetect4ML}, its outputs, and its specifications. Annotators relied on Zhang et al.'s ML code smell definitions~\cite{Zhang2022code} and on the released annotation material~\cite{SpecDetect4MLReplication}. Ground truth is defined at the occurrence level, with source locations recorded for accepted occurrences. For non-local occurrences spanning multiple files or statements, the benchmark records one canonical anchor location for counting and matching, while supporting cross-file evidence is documented in the annotation material. The annotation kit specifies the occurrence unit, inclusion and exclusion criteria, static boundary, and required evidence for each ML code smell, including whether the decisive evidence is local, non-local, configuration-dependent, or runtime-only.

The six annotators comprised one M.Sc.\ student, two Ph.D.\ students, and three industry experts with five to six years of experience in applied ML engineering and ML pipeline code review. The labelling unit was a smell occurrence identified within a selected system. Each annotator independently inspected the assigned systems and recorded all candidate occurrences according to the smell-specific guidelines. The resulting annotations were then compared across annotators at the occurrence level. Borderline cases, low-confidence annotations, and cases for which annotators reported different decisions were systematically re-examined and resolved through adjudication. Overall, 23 occurrences in the precision track and 147 occurrences in the recall track were flagged as uncertain by at least one annotator before adjudication. Before this step, Fleiss' $\kappa = 0.78$ was computed on pre-adjudication smell-occurrence labels grouped by file, canonical line, and context range after reviewer de-duplication.

To align with prior work, we also incorporate the \textit{CodeSmile} dataset~\cite{recupito2025_codesmile_lifecycle}, which provides adjudicated labels for 12 ML code smells on a stratified sample of 100 files. Our unified ground truth combines the adjudicated \textit{CodeSmile} subset with our constructed precision and recall sets, yielding coverage for all 22 ML code smells while retaining a benchmark-aligned anchor subset. Paper-level detection metrics cover R1 to R22 only, while R23 and R24 are excluded from detection evaluation and used solely as specification-authoring artefacts. The replication package provides the scripts used to recompute the supported-smell intersections and the CodeSmile-only subset.

To contextualise manual effort, validation covers 886 inspected alerts across 370 files and 58 systems for \textit{SpecDetect4ML}, 468 inspected alerts across 189 files and 43 systems for \textit{mlpylint}, and 1,234 inspected alerts across 158 files and 40 systems for \textit{CodeSmile}. The recall benchmark covers 605 files in the complete ML-relevant source subset and 2,547 adjudicated occurrence-level decisions extracted from 3,855 Python files across the 10 selected systems. LLM tasks are generated only for file/smell pairs where the smell is applicable to the file's library and pipeline context, yielding 1,125 prompt-based tasks rather than all 605$\times$22 possible pairs.

As a sensitivity check, we also recomputed recall after removing each selected system in turn. The resulting values remained within [86.34, 92.87], suggesting that the reported recall is not driven by a single annotated system. The full leave-one-system-out results are included in the replication package.

\subsection{Evaluation Metrics}
\label{subsec:metrics}

We compute all metrics at the smell occurrence level. Precision is estimated from the smell-balanced audit of analyser alerts. Let $S_P$ be the inspected alert sample, with $V$ the valid alerts and $V^c$ the invalid alerts. Precision is defined as $\hat{P} = |V|/(|V| + |V^c|)$.

Recall is computed on the independently annotated system-level benchmark. Let $G_R$ be the adjudicated ground truth occurrences and $A_R$ the analyser alerts on the same benchmark. An occurrence is recovered when an alert matches its smell type and canonical source location. Recall is defined as $\hat{R} = |A_R \cap G_R|/|G_R|$.

We report \Fonehat{} as the harmonic mean of these two estimates, with \Fonehat{} $= 2 \cdot \hat{P} \cdot \hat{R}/(\hat{P} + \hat{R})$. Because precision and recall come from complementary validation tracks, \Fonehat{} is used only as a compact summary, while precision and recall are interpreted separately. Aggregate \Fonehat{} intervals use a non-parametric track-wise bootstrap over inspected precision decisions and adjudicated recall occurrences.

\subsection{Results}
\label{sub:results}
In this section, we answer the research questions.

\rqheading{RQ$_1$: What portion of the ML code smell catalogue is statically observable at scale, and how are the resulting occurrences distributed across smells, files, and systems?}
\label{sub:RQ1}

Across 890 systems, \textit{SpecDetect4ML} detected 134,460 ML code smell occurrences across 34,209 affected files (Table~\ref{tab:rules_summary}). Under the detectability taxonomy introduced in Section~\ref{subsec:taxonomy}, these detections are concentrated in T1 to T4. These occurrence classes have decisive evidence that remains statically observable in local, intra-file, same-file interprocedural, or cross-file code structure. The absence of T5 and T6 counts should therefore be interpreted as part of the measurement scope, not as evidence that configuration-dependent or runtime-only ML code smell occurrences are absent from the corpus. When decisive evidence depends on external configuration or runtime behaviour, a static analyser can report it only if that evidence is materialised in the inspected source code.
Detection volume is highly uneven. The three most frequent ML code smells, R2, R6, and R4, account for 71.19\% of all detected occurrences. The distribution is also strongly skewed at the system level: 66 systems (7.42\%) have no detected occurrences, while a minority of systems concentrates a large share of the reports (Gini = 0.833). Thus, the 92.58\% of systems with at least one detected occurrence indicates widespread review candidates, not a direct defect rate. The intended use is prioritised review, not automatic remediation of all reports. Rare ML code smells also require cautious interpretation: R15 appears only seven times, so its validation results are best read as evidence of correct detection on the inspected cases rather than as a population-level estimate. By effect category, correctness-related ML code smells dominate the corpus with 85.84\% of all detections, partly because many correctness smells in this catalogue expose API-level or ordering evidence, whereas some efficiency and maintainability concerns require intent or runtime context and are harder to recover statically.

\begin{table}[!htb]
\centering
\scriptsize
\setlength{\tabcolsep}{2pt}
\renewcommand{\arraystretch}{0.8}
\caption{Detected ML code smell occurrences by smell, grouped by effect \cite{Zhang2022code}, and sorted by occurrence count.}
\label{tab:rules_summary}
\resizebox{0.87\linewidth}{!}{%
\begin{tabular}{@{}lrrrr@{}}
\toprule
\textbf{Rule ID} & \textbf{Occurrences} & \textbf{Occurrences (\%)} & \textbf{Affected Systems} & \textbf{Median/System} \\
\midrule
\rowcolor{gray!20}\multicolumn{5}{c}{\textit{\textbf{Correctness}}}\\[-2pt]
\midrule
R2  & 72,246 & 53.73 & 752 & 15 \\
R6  & 15,757 & 11.72 & 311 & 10 \\
R4  & 7,725  & 5.75  & 315 & 5 \\
R12 & 7,566  & 5.63  & 301 & 4 \\
R9  & 5,710  & 4.25  & 139 & 4 \\
R8  & 2,725  & 2.03  & 149 & 7 \\
R7  & 1,957  & 1.46  & 101 & 5 \\
R22 & 1,353  & 1.01  & 116 & 2 \\
R11 & 140    & 0.10  & 23  & 3 \\
R13 & 131    & 0.10  & 40  & 2 \\
R18 & 69     & 0.05  & 27  & 2 \\
R19 & 37     & 0.03  & 23  & 1 \\
\midrule
\rowcolor{gray!20}\multicolumn{5}{c}{\textit{\textbf{Efficiency}}}\\[-2pt]
\midrule
R10 & 6,430 & 4.78 & 456 & 3 \\
R1  & 1,973 & 1.47 & 98  & 4 \\
R17 & 144   & 0.11 & 54  & 1 \\
R3  & 62    & 0.05 & 10  & 4 \\
\midrule
\rowcolor{gray!20}\multicolumn{5}{c}{\textit{\textbf{Maintainability}}}\\[-2pt]
\midrule
R5  & 5,534 & 4.12 & 315 & 4 \\
R21 & 2,727 & 2.03 & 252 & 4 \\
R14 & 1,045 & 0.78 & 145 & 3 \\
R16 & 637   & 0.47 & 160 & 2 \\
R20 & 485   & 0.36 & 90  & 3 \\
R15 & 7     & 0.01 & 5   & 1 \\
\bottomrule
\end{tabular}%
}
\end{table}

\rqheading{RQ$_2$: How much does project-level CPG reasoning improve detection over AST-only analysis under identical ML code smell specifications and validation protocol?}
\label{sec:rq2}

We isolate the contribution of CPG-based project-level reasoning by executing \textit{SpecDetect4ML} in two modes under the same validation protocol. In AST-only mode, the analysis is restricted to local AST evidence, lexical statement order, and file-local syntax. It does not use resolved imports, call graphs, or def-use links. In CPG mode, the same ML code smell specifications are evaluated with additional control-flow, data-flow, same-file interprocedural, and cross-file relations. This controlled comparison keeps the ML code smell specifications fixed and varies only the program representation available to the analyser.

\begin{table}[h]
\footnotesize\centering
\renewcommand{\arraystretch}{0.85}
\setlength{\tabcolsep}{0.8pt}
\caption{Effectiveness comparison (AST-only vs CPG). Precision is estimated from the precision audit, and recall is estimated from the independent recall benchmark.}
\label{tab:rq2-summary}
\begin{tabular}{lccc}
\hline
\textbf{Mode} & \textbf{Precision (\%)} & \textbf{Recall (\%)} & \textbf{\Fonehat{} (\%)} \\
\hline
\textbf{AST-only} & \neu{90.43} & \negcell{68.62} & \negcell{78.04} \\
\textbf{CPG} & \neu{90.32} & \pos{88.14} & \pos{89.22}  \\
\hline
\end{tabular}
\end{table}

\noindent\textbf{Overall gain.}
Table~\ref{tab:rq2-summary} shows a clear gain from CPG reasoning. Precision stays comparable, from \textbf{90.43\%} to \textbf{90.32\%}, while recall rises from \textbf{68.62\%} to \textbf{88.14\%}. CPG therefore recovers many more ground-truth occurrences at similar precision. This gain is not driven by one high-volume smell: FNs show 20 decreases, 0 increases, and 2 ties (R3 and R19). An exact two-sided sign test over the 20 non-tied per-smell FN deltas gives $p=1.91\times10^{-6}$. Rows with very small support are interpreted descriptively.

\noindent\textbf{Smell-level interpretation.}
Table~\ref{tab:tp-fp-fn-by-smell} reports, for each smell, the ground-truth count (GT), the evaluation metrics for both modes (TP, FP, FN), and the deltas computed as $\Delta = \text{CPG} - \text{AST}$. We summarise net progress using $Err = FP + FN$ and \textit{Relative Error Reduction} $(RedErr)=100 \times \frac{Err_{AST}-Err_{CPG}}{Err_{AST}}$, which increases only when the CPG mode reduces the combined mass of FP and FN.

\begin{table}[h]
\centering
\footnotesize
\renewcommand{\arraystretch}{0.6}
\setlength{\tabcolsep}{3pt}
\setlength{\extrarowheight}{0pt}
\caption{Recall-benchmark diagnostic by smell under AST-only and CPG modes. FP counts are from the recall benchmark and are not the paper-level precision sample.}
\label{tab:tp-fp-fn-by-smell}
\resizebox{\columnwidth}{!}{%
\begin{tabular}{lrrrrrr}
\hline
\textbf{Smell} & \textbf{GT} & \textbf{AST-only $(TP,FP,FN)$} & \textbf{CPG $(TP,FP,FN)$} & \textbf{$\Delta$FP} & \textbf{$\Delta$FN} & \textbf{RedErr (\%)} \\
\hline
\textbf{R1}  & 22   & (15,1,7)   & (17,6,5)    & \negcell{5} & \pos{-2}   & \negcell{-37.5\%} \\
\hline
\textbf{R2}  & 1242 & (854,2,388)& (1058,6,184)& \negcell{4} & \pos{-204} & \pos{51.3\%} \\
\hline
\textbf{R3}  & 2    & (2,2,0)    & (2,6,0)     & \negcell{4} & \neu{0}    & \negcell{-200.0\%} \\
\hline
\textbf{R4}  & 123  & (73,3,50)  & (109,6,14)  & \negcell{3} & \pos{-36}  & \pos{62.3\%} \\
\hline
\textbf{R5}  & 105  & (73,1,32)  & (92,6,13)   & \negcell{5} & \pos{-19}  & \pos{42.4\%} \\
\hline
\textbf{R6}  & 344  & (246,0,98) & (299,6,45)  & \negcell{6} & \pos{-53}  & \pos{48.0\%} \\
\hline
\textbf{R7}  & 19   & (13,2,6)   & (19,2,0)    & \neu{0}     & \pos{-6}   & \pos{75.0\%} \\
\hline
\textbf{R8}  & 67   & (46,3,21)  & (64,4,3)    & \negcell{1} & \pos{-18}  & \pos{70.8\%} \\
\hline
\textbf{R9}  & 63   & (44,2,19)  & (60,3,3)    & \negcell{1} & \pos{-16}  & \pos{71.4\%} \\
\hline
\textbf{R10} & 137  & (95,0,42)  & (129,5,8)   & \negcell{5} & \pos{-34}  & \pos{69.0\%} \\
\hline
\textbf{R11} & 4    & (2,3,2)    & (3,6,1)     & \negcell{3} & \pos{-1}   & \negcell{-40.0\%} \\
\hline
\textbf{R12} & 76   & (53,8,23)  & (73,2,3)    & \pos{-6}    & \pos{-20}  & \pos{83.9\%} \\
\hline
\textbf{R13} & 16   & (11,4,5)   & (14,2,2)    & \pos{-2}    & \pos{-3}   & \pos{55.6\%} \\
\hline
\textbf{R14} & 91   & (66,5,25)  & (88,1,3)    & \pos{-4}    & \pos{-22}  & \pos{86.7\%} \\
\hline
\textbf{R15} & 10   & (6,2,4)    & (10,0,0)    & \pos{-2}    & \pos{-4}   & \pos{100.0\%} \\
\hline
\textbf{R16} & 87   & (55,5,32)  & (81,1,6)    & \pos{-4}    & \pos{-26}  & \pos{81.1\%} \\
\hline
\textbf{R17} & 10   & (7,4,3)    & (10,6,0)    & \negcell{2} & \pos{-3}   & \pos{14.3\%} \\
\hline
\textbf{R18} & 9    & (6,3,3)    & (9,0,0)     & \pos{-3}    & \pos{-3}   & \pos{100.0\%} \\
\hline
\textbf{R19} & 2    & (2,0,0)    & (2,0,0)     & \neu{0}     & \neu{0}    & \neu{NA} \\
\hline
\textbf{R20} & 35   & (24,4,11)  & (32,2,3)    & \pos{-2}    & \pos{-8}   & \pos{66.7\%} \\
\hline
\textbf{R21} & 74   & (51,2,23)  & (69,2,5)    & \neu{0}     & \pos{-18}  & \pos{72.0\%} \\
\hline
\textbf{R22} & 9    & (4,2,5)    & (5,6,4)     & \negcell{4} & \pos{-1}   & \negcell{-42.9\%} \\
\hline
\textbf{Total} & \textbf{2547} & \textbf{(1748,58,799)} & \textbf{(2245,78,302)} & \negcell{20} & \pos{-497} & \pos{55.7\%} \\
\hline
\end{tabular}%
}
\end{table}

\noindent\textbf{Where the gain comes from.}
Table~\ref{tab:tp-fp-fn-by-smell} shows that the main benefit of CPG reasoning is a substantial reduction in FNs. The total row reconciles the per-smell counts with the 2,547-occurrence recall benchmark: FN drops from 799 under AST-only analysis to 302 under CPG analysis, recovering 497 validated occurrences missed by AST-only mode, or 19.5\% of all validated occurrences. To expose how this gain maps to the taxonomy, we added post-hoc T1--T6 labels at the same occurrence unit used for recall. Labels describe where the decisive evidence resides, not which mode detected the occurrence, so the table should be read as an interpretive audit of the recall benchmark rather than as an independent causal proof.

\begin{table}[t]
\centering
\scriptsize
\renewcommand{\arraystretch}{1}
\caption{Detected occurrences by detectability taxonomy.}
\label{tab:taxonomy-detection}

\begin{tabular}{l|rrr}
\toprule
\textbf{Class} & \textbf{Ground Truth} & \textbf{CPG} & \textbf{AST-only} \\
\midrule
T1 & 383  & \neu{353}  & \neu{353}  \\
T2 & 1221 & \neu{1173} & \neu{1173} \\
T3 & 607  & \pos{594}  & \negcell{205}  \\
T4 & 94   & \pos{81}   & \negcell{17}   \\
T5 & 132  & \pos{44}   & \negcell{0}    \\
T6 & 110  & \neu{0}    & \neu{0}    \\
\midrule
\textbf{Total} & \textbf{2547} & \pos{2245} & \negcell{1748} \\
\bottomrule
\end{tabular}
\end{table}

Table~\ref{tab:taxonomy-detection} makes the taxonomy-level pattern explicit. AST-only analysis is already effective on occurrences whose evidence remains local or intra-file: it recovers most T1 and T2 cases, where the relevant API call, omitted argument, or statement ordering pattern is visible in the parsed source structure. Its limitations become clearer on non-local occurrences. For T3 cases, where the decisive evidence is distributed across functions in the same file, AST-only recovers only a minority of the benchmark occurrences, whereas CPG mode recovers 594 of 607 cases. For T4 cases, where evidence is split across files and requires imports, aliases, wrappers, or calls to be resolved, CPG mode also improves substantially, recovering 81 of 94 cases compared with 17 for AST-only analysis.

This pattern clarifies what CPG reasoning adds. The gain does not come from changing the ML code smell definitions, since the DSL specifications are held fixed. Instead, CPG mode exposes source-level relations that local AST matching cannot systematically recover: control-flow ordering, def-use dependencies, same-file interprocedural links, and project-level import or wrapper resolution. This is why the main recall gain appears in T3 and T4 rather than in clearly local T1 cases. T5 remains only partially recoverable, because configuration-dependent evidence is detected only when configuration values are materialised in the inspected source code. T6 remains outside the scope of deterministic static analysis because the decisive evidence depends on runtime behaviour.

Overall, the taxonomy supports the main claim of RQ2: CPG-based project-level reasoning expands recall mainly for statically observable non-local occurrences, while configuration-dependent and runtime-only cases remain limits for source-only static analysis.

\noindent\textbf{False positives under CPG reasoning.}
The recall gain comes with additional FP in some smells. Most rows still improve in combined error because the FN reduction dominates, but R1, R3, R11, and R22 show the boundary condition: broader propagation can over-generalise when structurally similar patterns do not reflect the same engineering intent. For R1 \textit{Broadcasting Feature Not Used}, all six CPG-mode false positives in the precision sample occur in TensorFlow tests around \texttt{tile}, where recovered paths are negative tests or edge-case checks rather than missed broadcasting opportunities. Another example is R9 \textit{Gradients Not Cleared}, where a recovered occurrence may look like a missing \texttt{zero\_grad()} call while the code actually implements deliberate gradient accumulation.

\rqheading{RQ$_3$: How effective is \textit{SpecDetect4ML}, in precision, recall, and \Fonehat{}, compared with published static analysers for ML code smells and normalised LLM baselines?}
\label{sec:rq3}
\newcommand{\pCI}[3]{\shortstack{#1\\{\footnotesize[#2, #3]}}}

\begin{table}[t]
  \centering
  \caption{Detection performance on the ground truth. Rows with different coverage are evaluated on supported-smell intersections.}
  \label{tab:rq3-global}
  \resizebox{1\columnwidth}{!}{%
    \begin{tabular}{l|ll|rrr}
      \toprule
      \textbf{Analyser} & \textbf{Approach} & \textbf{Coverage (smells)} & \textbf{$P$ (95\% CI) (\%)} & \textbf{$R$ (95\% CI) (\%)} & \textbf{\Fonehat{} (95\% CI) (\%)} \\
      \midrule
      \textit{\textbf{SpecDetect4ML}} & DSL + CPG analysis &
      \pos{22}  &
      \pCI{90.32}{88.09}{92.18} &
      \pos{\pCI{88.14}{86.83}{89.34}} &
      \pos{\pCI{89.22}{87.45}{90.74}}
      \\
      \hline
      \textbf{CodeSmile} & AST-based analyser &
      \negcell{12} &
      \pCI{90.96}{89.12}{92.51} &
      \pCI{61.27}{57.50}{64.91} &
      \pCI{73.22}{69.92}{76.25}
      \\
      \hline
      \textbf{Qwen 3 30B Instruct} & LLM baseline &
      \pos{22} &
      \pCI{91.50}{89.90}{92.87} &
      \pCI{48.22}{46.30}{50.15} &
      \pCI{63.16}{61.37}{64.94}
      \\
      \hline
      \textbf{Claude Sonnet 4.6} & LLM baseline &
      \pos{22} &
      \pCI{92.17}{90.40}{93.64} &
      \pCI{37.72}{35.87}{39.61} &
      \pCI{53.53}{51.61}{55.41}
      \\
      \hline
      \textbf{DeepSeek V4 Flash} & LLM baseline &
      \pos{22} &
      \negcell{\pCI{87.59}{85.21}{89.63}} &
      \pCI{28.88}{27.17}{30.66} &
      \pCI{43.44}{41.32}{45.50}
      \\
      \hline
      \textbf{mlpylint} & AST-based analyser &
      20 &
      \pos{\pCI{92.31}{89.53}{94.39}} &
      \negcell{\pCI{6.38}{5.50}{7.41}} &
      \negcell{\pCI{11.95}{10.36}{13.74}}
      \\
      \bottomrule
    \end{tabular}%
  }
\end{table}

\noindent Table~\ref{tab:rq3-global} shows that \textit{SpecDetect4ML} achieves the strongest overall balance between precision and recall, with 90.32\% precision, 88.14\% recall, and 89.22\% \Fonehat{} on the unified ground truth. These values are estimates over adjudicated validation samples, not corpus-wide defect rates. The published AST-based analysers remain competitive in precision, with CodeSmile reaching 90.96\% and mlpylint 92.31\%, but their recall is substantially lower. CodeSmile reaches 61.27\% recall on its 12-smell supported intersection, while mlpylint reaches only 6.38\% recall on its 20-smell intersection. This indicates that existing AST-based analysers can be reliable when they emit alerts, but their coverage remains limited, especially when occurrences require non-local reasoning.
The LLM baselines show the opposite trade-off. Their precision is comparable to the static analysers, especially for Qwen 3 30B Instruct and Claude Sonnet 4.6, but their recall remains far below \textit{SpecDetect4ML}. Qwen reaches 48.22\% recall, Claude reaches 37.72\%, and DeepSeek V4 Flash reaches 28.88\%. Thus, prompt-based detection improves substantially over mlpylint in recall, but it does not match the systematic coverage of the CPG-based analyser. Read together with the RQ2 taxonomy analysis in Table~\ref{tab:taxonomy-detection}, these results suggest that local explicit occurrences are accessible to several approaches, whereas non-local and omission-style occurrences benefit from deterministic project-level static reasoning.
To ensure comparability, the values in Table~\ref{tab:rq3-global} are not copied from the original tool papers. Each analyser or prompt-based detector is rerun under our shared occurrence-level protocol and normalised to the same detection unit: ML code smell identifier, file, line, and source region. Comparisons are restricted to supported-smell intersections when baselines cover fewer than 22 ML code smells, so the table should be read together with the coverage column. This matters especially for mlpylint, which is precise but sparse on its 20-smell intersection ($P$=92.31\%, $R$=6.38\%). Table~\ref{tab:tp-fp-fn-by-smell} is a recall-benchmark diagnostic table, so its TP/FP/FN counts are not expected to reproduce the precision or aggregate \Fonehat{} values in Table~\ref{tab:rq3-global}.

\noindent\textbf{Per-smell view.}
Per-smell \Fonehat{} scores are provided in the replication package. \textit{SpecDetect4ML} is more uniform across the catalogue, whereas external static analysers and LLM baselines show sparser profiles due to limited coverage or less systematic recovery for some smells. We interpret low-support smells descriptively rather than as stable per-smell estimates.
The per-smell profile also helps interpret the aggregate scores. \textit{SpecDetect4ML} maintains high \Fonehat{} on both frequent local smells and several non-local smells, while CodeSmile and mlpylint have sharper peaks on the subset of smells they explicitly support. The LLM baselines show a different shape: they are stronger on local explicit patterns, but their profile contracts on omission-style, lifecycle-dependent, or cross-file smells. This reinforces the RQ2 result that the main advantage of CPG reasoning is breadth across statically observable non-local cases, not only a higher aggregate score.

\noindent\textbf{LLM baseline interpretation.}
The LLM comparison uses a one-shot, temperature-0, JSON-schema protocol over bounded source snippets, without iterative repair, self-consistency, retrieval, tool use, or full-repository context. We therefore interpret it as a conservative prompt-based baseline, not as a claim about the full design space of LLM-assisted or agentic analysis. Runtime also differs: Qwen, Claude, and DeepSeek require 1{,}810.5~s, 3{,}510.8~s, and 5{,}152.3~s respectively, versus 884.6~s for CPG mode and 283.5~s for AST-only mode. LLMs do better on local explicit patterns such as R1, R4, R9, R12, R20, and R22, but are less stable for R6, R10, R11, R16, and R19, where evidence is missing, non-local, lifecycle-dependent, or rare.

\rqheading{RQ$_4$: What do representative true positives, false positives, and false negatives reveal about statically observable, configuration-dependent, and runtime-only ML code smell occurrences?}
\label{sec:rq4}

Representative cases show both the value and the boundary of static reasoning. CPG-only TPs are often T3/T4 cases where decisive evidence is distributed across functions or files, as in Listing~\ref{lst:R11_cpg_recovered}, or in R4 \textit{Improper Train/Eval Mode Toggling}, where mode switches and evaluation calls are separated by helper abstractions. High scores for smells such as R6 or R11 should still be read within this scope: they show recovery when deterministic options, preprocessing-before-split patterns, or related lifecycle evidence are visible in source code, not full semantic judgement over all ML behaviour.

The CPG-only cases are useful because they show a recurring pattern rather than isolated successes. For example, wrappers often hide whether a dataset split, preprocessing call, seed, or mode switch belongs to the same pipeline stage. Local AST matching can observe each call, but it cannot reliably connect the caller, callee, imported alias, and later consumer. CPG reasoning makes these links explicit enough to recover review candidates while still preserving the original smell definitions.

FPs arise when a suspicious structure remains compatible with valid intent, such as R9 cases where a missing \texttt{zero\_grad()} reflects deliberate gradient accumulation, or test code that intentionally exercises edge cases. These examples are not random noise. They mark the boundary between structural evidence and engineering intent. We therefore treat reports as review candidates rather than automatic defects. A useful alert is one that points reviewers to a plausible risk with enough source context to accept or dismiss it.

Remaining FNs are dominated by evidence outside source-only static reach. Among the 302 CPG-mode FNs, the reviewed labels assign 30 to T1, 48 to T2, 13 to T3, 13 to T4, 88 to T5, and 110 to T6. Thus 198 FNs (65.6\%) require configuration-dependent or runtime-only evidence. Low-confidence recall cases are mostly T1/T2, indicating that annotation uncertainty is concentrated in statically visible but intent-sensitive cases. High-volume rules such as R2, R4, and R6 indicate where review effort may be needed, not where automatic remediation is always appropriate.

\rqheading{RQ$_5$: To what extent does \textit{SpecDetect4ML} scale to a large corpus of ML-based systems, and what overhead does project-level CPG reasoning introduce over AST-only analysis?}
\label{sec:rq5}

We assess scalability on 890 GitHub ML systems, totalling 142{,}747 Python files and 20.5M LOC, on an Apple M4 Pro with 24~GB RAM. Across the full corpus, \textit{SpecDetect4ML} completes end-to-end analysis in 26{,}246.1~s (7h 29min), or 1.28~s per 1{,}000 LOC. Runtime is right-skewed. The median is 3.78~s per system, 90\% complete within 6.20~s, and 99\% within 14.5~s. The largest repositories account for a disproportionate share of total runtime.

\begin{center}
\scriptsize
\renewcommand{\arraystretch}{0.75}
{\captionsetup{hypcap=false}%
\captionof{table}{General statistics of the analysis.}\label{tab:general_stats}}
\resizebox{0.43\columnwidth}{!}
{%
\begin{tabular}{@{}lr@{}}
\toprule
\textbf{Metric} & \textbf{Value} \\ \midrule
Number of GitHub systems analysed & 890 \\
Size of all systems & 29.44~GB \\
Number of \texttt{.py} files & 142{,}747 \\
Lines of code (LOC) & 20{,}454{,}021 \\
Files containing at least one code smell & 34{,}209 \\
Total number of code smells detected & 134{,}460 \\
Total analysis time & 26{,}246.1 s (7h 29min) \\
Average analysis time per system & 29.49 s \\
Median analysis time per system & 3.78 s \\
90$^{\text{th}}$-percentile time per system & 6.20 s \\
99$^{\text{th}}$-percentile time per system & 14.5 s \\
Aggregate time / 1{,}000 LOC & 1.28 s \\
\bottomrule
\end{tabular}}%
\end{center}

This scalability has measurable cost: on the AST-only/CPG comparison subset, median runtime rises from 283.5~s to 884.6~s (3.12$\times$), and median time per file from 0.094~s to 0.203~s (2.16$\times$). In practice, the analyser is therefore best used for warning-oriented review and targeted audits, while T5/T6 cases motivate complementary configuration-aware or dynamic checks.

The overhead is acceptable for batch auditing, but it is not free. The largest repositories dominate total analysis time because project-level resolution must build and traverse more relations before a rule can query them. This suggests two practical deployment modes: lightweight AST-only screening when fast local feedback is preferred, and CPG mode for pre-release audits, migration reviews, or repository-wide quality studies where recall on non-local occurrences is more important.

\subsection{Threats to Validity}
\label{subsec:threats}

\textbf{Construct validity.} SpecDetect4ML operationalises smells as statically observable Python source patterns. This matches the paper's goal, but it can miss evidence available only through external configuration, framework defaults, data-dependent behaviour, or runtime execution. We mitigate boundary ambiguity through occurrence units, smell-specific guidelines, the T1--T6 taxonomy, and a review-candidate interpretation of alerts. The taxonomy is therefore not used to claim that static analysis can decide every smell. It records where the decisive evidence resides.

\textbf{Internal validity.} Manual labels may depend on framework knowledge and developer intent, for example in gradient-accumulation cases or tests that intentionally violate production conventions. We mitigate this with guidelines, six ML-experienced annotators, adjudication, agreement reporting, and non-author annotators blind to SpecDetect4ML. LLM baselines may also be affected by nondeterminism, provider changes, prompt sensitivity, and open-source data leakage. We release prompts, raw outputs, model identifiers, access dates, and normalisation scripts so that the comparison can be re-executed.

\textbf{External and conclusion validity.} The corpus targets open-source Python ML systems, and recall is estimated on a 10-system benchmark rather than an exhaustive 890-system ground truth. The baseline comparisons also use supported-smell intersections when tools cover different parts of the catalogue. We therefore report confidence intervals, separate precision and recall tracks, and interpret per-smell results as explanatory rather than universal.

\section{Related Work}
\label{sec:related_work}

\noindent ML-based systems combine source code, data, configuration, and model behaviour, creating quality issues that traditional software analyses only partially address~\cite{AsynthesisOfGreen2024,Shaw2022,passi2018,wei2022APIRecommendation}. ML code smells provide source-level indicators of such issues. The main reference is Zhang et al.'s catalogue of 22 ML code smells~\cite{Zhang2022code}. Existing tools include CodeSmile~\cite{recupito2025_codesmile_lifecycle} and \textit{mlpylint}~\cite{Hamfelt2023MLpylint}. They work well for local patterns but become less reliable when relevant information is spread across wrappers, imports, execution order, or data flows.

\noindent This limitation motivates separating smell definitions from detection mechanisms. DSL research advocates separating domain concepts from implementation details~\cite{DSL}. \textit{SpecDetect4ML} follows this principle by decoupling ML smell specifications from the analysis backend. Our approach is related to CodeQL~\cite{codeql} and Semgrep~\cite{semgrep}, but packages recurrent framework and program-structure logic as reusable ML-oriented predicates and executes the same specifications with AST-only and CPG analysis. This is the core novelty of our work: the specifications stay fixed while the backend changes what is recoverable.

\noindent Complementary work on ML program analysis studies framework-specific API misuse rather than a reusable smell catalogue. Ariadne~\cite{Dolby2018Ariadne} and recent work on deep-learning API misuses~\cite{wei2024Demystifying} focus on API-level evidence and framework-specific incorrectness patterns. \textit{SpecDetect4ML} instead targets a broader catalogue and asks which occurrences are recoverable when the same specifications are evaluated with different static backends.

\noindent LLM-assisted approaches occupy another complementary point in this space. They can reason from natural-language smell descriptions and code snippets, which is useful for local or explicit patterns, but their outputs must still be normalised to occurrence-level locations before they can support precision and recall claims. Our evaluation therefore treats LLMs as prompt-based baselines under a shared matching protocol, rather than as replacements for deterministic static reasoning.

\noindent The backend design is motivated by graph-based program analysis. CPGs integrate AST, CFG, and PDG information~\cite{Yamaguchi2014CPG} and are widely used in software security because they capture syntax, control flow, and data dependencies across abstractions and indirections~\cite{Backes2017PHPVulnCPG,Yamaguchi2014CPG,JoernDocs,CPGSpec}. These properties make them attractive for ML code smell detection whenever decisive evidence is distributed across wrappers, aliases, execution order, or cross-file data movement. Our contribution is not the CPG abstraction itself, but its adaptation and empirical isolation for ML code smell detection. Related API-aware analysers, WALA-based analyses, dynamic frameworks such as DynaPyt~\cite{Eghbali2022DynaPyt}, and LLM-assisted smell detectors such as iSMELL~\cite{ismell} occupy complementary points in the design space. Dynamic analysis can observe behaviours that static analysis cannot, but it depends on executable workloads and observed paths. In contrast, \textit{SpecDetect4ML} targets reusable smell specifications and deterministic static project-level reasoning for statically observable non-local occurrences.

\section{Conclusion}
\label{sec:conclusion}

We presented \textit{SpecDetect4ML}, a static analyser that operationalised 22 ML code smells using CPG views with project-level resolution. We evaluated it on 890 Python ML-based systems comprising more than 20M LOC and a system-level recall benchmark over 10 selected systems. Under identical ML code smell specifications, CPG-based reasoning raised recall from 68.62\% to 88.14\% compared with AST-only analysis, while keeping CPG precision comparable at 90.32\%. These results showed that project-level static reasoning expanded the detectable portion of non-local ML code smell occurrences, while configuration-dependent and runtime-only occurrences remained outside our source-only static claims.

\noindent\textbf{Data Availability.} The anonymous replication package~\cite{SpecDetect4MLReplication} includes the tool, executable smell specifications, corpus and commit metadata, sampling and annotation materials, adjudicated labels, normalised outputs for all evaluated tools and LLM baselines, prompts, model-access metadata, and scripts to reproduce the reported tables.

\bibliographystyle{IEEEtran}
\bibliography{bibli}

\end{document}